%% file: main.tex
\pdfoutput=1
\documentclass[sigconf, anonymous=false,authorversion,nonacm]{acmart}

\usepackage{subcaption}
\usepackage{color, colortbl, soul}

\begin{document}

\title{Crowdotic: A Privacy-Preserving Hospital Waiting Room Crowd Density Estimation with Non-speech Audio}



\author{Forsad Al Hossain}
\authornote{Both authors contributed equally}
\orcid{}
\affiliation{%
\institution{Manning College of Information \& Computer Sciences, University of Massachusetts Amherst}
\streetaddress{}
\city{Amherst}
\state{MA}
\postcode{01003}
\country{USA}}
\email{falhossain@cs.umass.edu}

\author{Tanjid Hasan Tonmoy}
\authornotemark[1]
\orcid{}
\affiliation{\institution{Halıcıoğlu Data Science Institute, University of California, San Diego}
\streetaddress{}
\city{San Diego}
\state{CA}
\postcode{92093}
\country{USA}}
\email{mtonmoy@ucsd.edu}

\author{Andrew A. Lover}
\affiliation{%
 \institution{School of Public Health \& Health Sciences, University of Massachusetts Amherst}
  \city{Amherst}
  \state{MA}
  \postcode{01003}
  \country{USA}}
\email{alover@umass.edu}

\author{George A. Corey}
\affiliation{%
  \institution{University Health Services, University of Massachusetts Amherst}
  \city{Amherst}
  \state{MA}
  \postcode{01003}
  \country{USA}}
\email{gcorey@uhs.umass.edu}

\author{Mohammad Arif Ul Alam}
\affiliation{%
  \institution{Miner School of Computer \& Information Sciences, University of Massachusetts Lowell}
  \city{Lowell}
  \state{MA}
  \postcode{01854}
  \country{USA}}
\email{MohammadArifUl_Alam@uml.edu}

\author{Tauhidur Rahman}
\affiliation{%
  \institution{Halıcıoğlu Data Science Institute, University of California, San Diego}
  \city{San Diego}
  \state{CA}
  \postcode{92093}
  \country{USA}}
\email{trahman@ucsd.edu}

\newcommand{\etal}{\textit{et al}.}
\renewcommand{\shortauthors}{Al Hossain et al.}

\begin{abstract}
Privacy-preserving crowd density analysis finds application across a wide range of scenarios, substantially enhancing smart building operation and management while upholding privacy expectations in various spaces. We propose a non-speech audio-based approach for crowd analytics, leveraging a transformer-based model. Our results demonstrate that non-speech audio alone can be used to conduct such analysis with remarkable accuracy. To the best of our knowledge, this is the first time when non-speech audio signals are proposed for predicting occupancy. As far as we know, there has been no other similar approach of its kind prior to this. To accomplish this, we deployed our sensor-based platform in the waiting room of a large hospital with IRB approval over a period of several months to capture non-speech audio and thermal images for the training and evaluation of our models. The proposed non-speech-based approach outperformed the thermal camera-based model and all other baselines. In addition to demonstrating superior performance without utilizing speech audio, we conduct further analysis using differential privacy techniques to provide additional privacy guarantees. Overall, our work demonstrates the viability of employing non-speech audio data for accurate occupancy estimation, while also ensuring the exclusion of speech-related content and providing robust privacy protections through differential privacy guarantees.
\end{abstract}

\keywords{occupancy estimation, neural networks, audio processing, machine learning}


\maketitle

\input{intro}
\input{related_work}
\input{dataset}

\input{overview}
\input{discussion}


\bibliographystyle{ACM-Reference-Format}
\bibliography{bibliography} 

\clearpage

\end{document}

%% file: intro.tex
\section{Introduction}
Estimating the number of people present within a specific area over a period of time has many applications. These include management of public spaces and events \cite{saleh2015recent}, surveillance and security \cite{rahmalan2006crowd}, preventing the spread of infectious diseases \cite{muthunagai2023crowd}. Quantification of crowd density empowers relevant authorities to make well-informed decisions, enabling smart building management to optimize safety, enhance comfort, minimize energy usage, and efficiently allocate resources for the management of diverse indoor and outdoor spaces. In addition to ensuring the accuracy and reliability, it is crucial to address the potential concerns related to the intrusiveness and privacy implications of the technology used for such crowd analytic system. To instill public trust and confidence in the smart building management framework, it is crucial to address potential privacy concerns, particularly in indoor spaces where privacy expectations are higher. Achieving such highly accurate and privacy preserving system requires the thoughtful selection of suitable sensing modalities and robust data processing methods, and minimizing the amount of information that is retained for future analysis. This emphasis on privacy-preserving practices contributes to the broader goal of integrating crowd analytics into smart buildings while respecting individuals' privacy rights.

Most of the current research and and deployed methods primarily utilize computer vision-based techniques for the task \cite{fiandeiro2023modernized, fan2022survey}. However, there remains significant concern regarding the use of different types of cameras and vision sensors, as they may fail to conform to the privacy expectation of users. Additionally, these techniques have several technical limitations such as susceptibility to occlusions, limited field of view and lighting conditions. Thus many research works investigated the use of other sensing modalities and their fusion for this task. These modalities include radio frequency (RF) \cite{rf-review-1}, acoustics \cite{perturb-audio-privacy}, environmental sensors (e.g. $\textit{CO}_{2}$, temperature, light, motion) \cite{6902229, chen2020deep}. However, all of them have been shown to be location/environment dependent ($\textit{CO}_{2}$, thermal, motion, vision, rf), with some also posing potential privacy concerns (vision, audio).

The acoustic signatures generated by crowds contain rich information that can be used to extract useful information, such as occupancy. However, one of the main challenges in using audio-based analysis is ensuring user privacy since the audio may include speech. In order to ensure privacy, it is necessary to record and analyze the audio in a way that makes it impossible to discern any speech content and prevents the identity of the speakers from being revealed. In this work, we deploy a sensor-based platform with an on-device machine model to filter speech and only capture non-speech audio. Our model is able to distinguish speech and non-speech audio with high accuracy \cite{al2020flusense} and we only use the non-speech signal for our analysis. We demonstrate that it is possible to reliably estimate crowd statistics in a coarse time scale using only non-speech, ensuring privacy. We use a microphone array to capture the audio, which captures acoustic information in multiple channels and angle of arrival information. Compared to vision-based estimation using a thermal camera, we obtain improved accuracy with non-speech audio.

We run a three-month-long study by deploying the platform to capture non-speech audio in the waiting room of a large hospital. We collected a large-scale dataset of non-speech audio in a realistic environment. We show that it is possible to estimate the occupancy with a high degree of accuracy using only non-speech audio. We compared the results from our non-speech-based models to the thermal camera-based model. Our results showed that the non-speech audio-based models outperformed the thermal-based model, especially for shorter time windows. We used a transformer-based model for the non-speech-based occupancy estimation. With this method, we have successfully designed an occupancy estimation system that harnesses the power of modern deep-learning techniques without requiring complex on-device deep learning models. This approach prioritizes privacy by excluding speech content and implementing differential privacy techniques. As a result, our system strikes a balance between accuracy, efficiency, and safeguarding sensitive information.

The main contributions in this paper are as follows-
\begin{enumerate}
    \item We propose a privacy-preserving non-speech audio-based crowd estimation approach and show that it is possible to obtain high levels of accuracy using only non-speech audio. To the best of our knowledge, this is the first-ever attempt to accomplish occupancy counting from non-verbal audio.
    \item We collected a large dataset of non-speech audio by deploying our sensor-based platform in the waiting room of a large hospital and show that our model works robustly with real-life audio data.
    \item We compare our non-speech approach with thermal camera-based occupancy estimation and other baseline approaches and obtain better results.
    \item In addition to using the privacy-preserving non-speech modality for analysis, we further enhance privacy preservation by employing differential privacy techniques.
\end{enumerate}

%% file: related_work.tex
\section{Related work}
Multiple applicable methods can be applied for the estimation of occupancy in a particular location. Traditional methods use a variety of sensing methods, including vision-based sensors, $\textit{CO}_{2}$ detection, EM waves, PIR sensors, sounds; including indirect methods such as computer activity or energy consumption. Some of the most popular methods are :

\subsection{$\textit{CO}_{2}$ based occupancy counting} As every occupant in a place exhales carbon-di-oxide, the concentration of carbon-di-oxide can be used as an indicator of occupancy. Although it showed highly accurate results for occupancy detection, to estimate occupancy, the results are in general poor.  To estimate occupancy either physics-based models \cite{physicsco21},  \cite{physicsco22}, \cite{physicsco23} or data-driven methods \cite{ubico2}   \cite{datamultimodalco21}, \cite{datamultimodalco22}, \cite{zhou2020novel} besides CO2 or even a hybrid approach utilizing both physics-based and data-based models \cite{physicsdataco2} are utilized.  Usually, physics-based models suffer from errors due to real-world complexities not accounted for in the ideal physics equation, including ventilation, exhaust fans, air conditioner, or frequency of door opening or closing. On the other hand, data-driven methods either perform poorly compared to physics-based models due to a lack of relevant data or require multiple sensor integration (thermal, light, humidity, etc.), which can be costly.
\subsection{RF signal based occupancy counting} Different electromagnetic spectrum bands with different properties are usually used for RF-based occupancy counting. Works in \cite{deviceocc1}, 
\cite{deviceocc2}, \cite{deviceocc3} tries to measure the occupancy based on the devices carried by the users. Several other works used device-free schemes to estimate occupancy from wifi signals, either using RSSI information \cite{wifirssi1} or by using CSI property \cite{wificsi1}, \cite{wificsi2}. However, these methods are not well suited for different environments and get easily affected by environmental factors.

Another popular method for people is to use an IR-UWB radar (\cite{uwb1}, \cite{uwb2}, \cite{uwb3}, \cite{uwb4}). This kind of system usually consists of one or multiple pairs of transmitter-receiver pairs. The transmitter sends an impulse, and the antenna receives the reflected signal.  The same signal gets reflected from different objects at different distances. The difference between two successive frames indicates movement, which can be indicative of human presence. However, as it detects movement, there is a good chance of getting interfered with by any movement, which complicates the detection and counting of humans.

\subsection{Sound-wave/audio-based occupancy estimation}

The usual audio-based occupancy estimation method utilizes speaker-diarization techniques (
\cite{spkaudio1}, \cite{spkaudio2}, \cite{spkaudio3}) to distinguish individual speakers to estimate the occupancy. However, the downside of such approaches is that it is assumed that everyone present in the room is talking, which might not be true in all settings, for example, say in a waiting room or a bus stop or in general, any public place where there is no social connection between different people occupying the area. Also, these techniques are privacy insensitive when the data needs to be sent to a backend to further processing.

Another active sound-based method for counting occupancy uses ultrasonic-chirps to detection motion from people (\cite{usonic1}, \cite{usonic2}). In this approach, a transmitter outputs a chirp, and a microphone listens for reverberation to detection. This technique is similar to IR-UWB based RF methods and, similar to them, also suffer from distinguishing real humans from other motion sources like air-conditioning or exhaust fans.

\subsection{Video-based method} Video-based occupancy measurements systems use cameras to count and track the number of people in indoor settings. These methods either utilizes RGB video stream \cite{rgbocc1}, \cite{rgbocc2}, \cite{rgbocc3} , \cite{rgbocc4}, \cite{rgbocc5}, \cite{rgbocc6}  or thermal camera based methods \cite{thermalocc1}, \cite{thermalocc2}. There are also other group of methods that measures occupancy by counting and tracking the people that crosses the entrance and exit of indoor locations [\cite{lineocc1}, \cite{lineocc2}, \cite{lineocc3}, \cite{lineocc4}, \cite{lineocc5}] to estimate occupancy of a location.

All vision-based methods, while being highly accurate when the camera field of view covers the whole location. However, they become unreliable when the location cannot be monitored with a single camera. In those settings, multiple cameras and algorithms to merge various video streams are needed to count the total number of people. So it becomes more and more complicated and costly to monitor space with compartments such as cubicles or walls. Also, a place might have multiple exits and entrance locations and monitoring each entrance exit requires multiple camera streams.  Besides that, camera-based occupancy counting is highly undesirable from a privacy perspective. Video contains numerous unnecessary pieces of information that can be used for action recognition or person identification. As the state-of-the-art methods for people counting from video require computation-heavy deep learning models counting the number of people, the data usually needs to go to a backend to effectively merge and compute the count of people from the video streams. Indeed, because of the above-mentioned concerns, there has been a movement to ban video-based surveillance \cite{facerecogban}

\subsection{Multimodal fusion-based method}
In addition to using single sensing modality, many works focus of fusing multiple sensing modalities. Thse approaches include fusion of environmental sensors such as \cite{en16052388, colace2022room}, and inclusion of image and audio modalities \cite{TAN2022111828}. These methods similar limitations associated with each sensing modality as discussed and also require higher overhead due to increased points of failure and need for synchronization.  

\subsection{Approaches for privacy preservation}
In addition to utilizing less intrusive sensing modalities such as PIR sensors \cite{stamatescu2021privacy}, approaches such as federated learning has been used to ensure privacy for crowd estimation tasks \cite{fed-learning-approach1}. Other approaches include perturbing audio data to prevent identification of speech and speaker \cite{perturb-audio-privacy}. However, these approaches have several limitations such as the use of simulated data that may not capture the challenging conditions associated with realistic data and fail to preserve privacy aspects such as speaker identity.

%% file: dataset.tex
\section{Dataset}
\subsection{Data collection platform}
\begin{figure}[!tbp]
  \centering
  \includegraphics[width=1\linewidth]{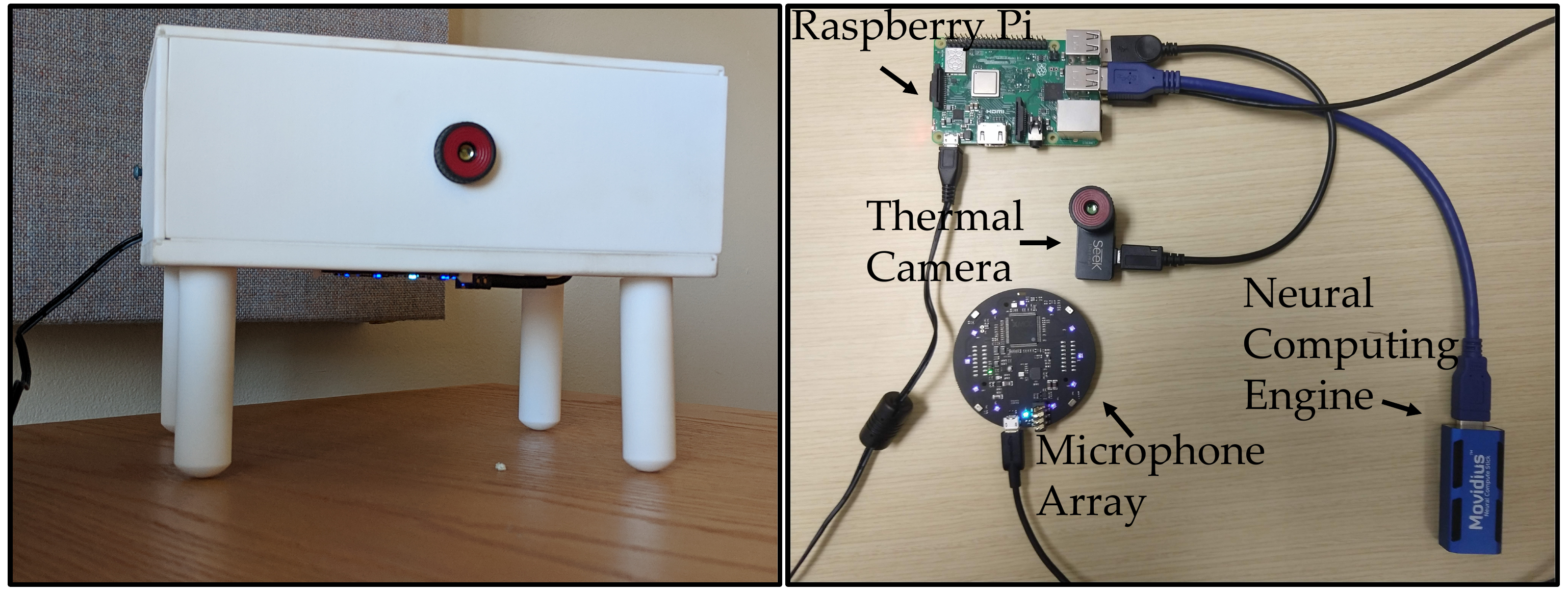}
  \caption{Front view and the components of the sensor platform.}
  \label{fig:flusense}
\end{figure}

We collected data for our non-speech dataset from an on-the-edge device consisting of a thermal camera, a microphone array and a neural computing stick attached to a Raspberry Pi platform. Following is a short description of the components of the platform, illustrated in Figure \ref{fig:flusense}:
\begin{itemize}
\item Microphone array: We used a ReSpeaker Microphone array \cite{ReSpeaker} to collect audio data. The ReSpeaker microphone array has four microphone arrays that collect audio streams. Then it produces a cleaner single-channel audio stream using on-device audio beamforming.
\item Thermal camera: A SeekCompact Pro \cite{SEEKThermal} thermal camera was used to take thermal images. It has a resolution of 320x240 pixels.
\item Neural Computing Stick: An Intel Neural Computing Stick \cite{intelncs} was used to accelerate deep learning computations on the device.
\item Raspberry pi: A Raspberry Pi was used to manage all the other sensors and store the data on a hard drive.
\end{itemize}

We have an on-device speech recognition model to eliminate any speech data from the audio stream. We built our speech-recognizer using TIMIT dataset \cite{garofolo1993timit} as a speech data source.  This dataset contains speech 630 American English speakers with all eight major dialects in the USA. Each speaker speaks eight major dialects. We also utilized a custom-labeled google audio set \cite{AudioSet} as a negative example source of speech. We use both of these datasets to develop a custom CNN-based speech-detection model. If our speech recognizer model determines that an audio snippet has a probability of being speech content greater than $0.5$, we do not store that snippet. We only retrain audio snippets with a speech probability less than 0.5, and we encrypt the data before storing.

All of our collected data were stored on a local hard drive after a two-stage encryption scheme. In the first stage, the original data gets encrypted with a randomly generated AES key. Then, in the second stage, that randomly-generated AES key was encrypted and stored with a public key. It is only possible to decrypt the stored AES key using a private key that is only available to the researchers.

\subsection{Deployment}

\begin{figure}[!htbp]
  \centering
  \includegraphics[width=1\linewidth]{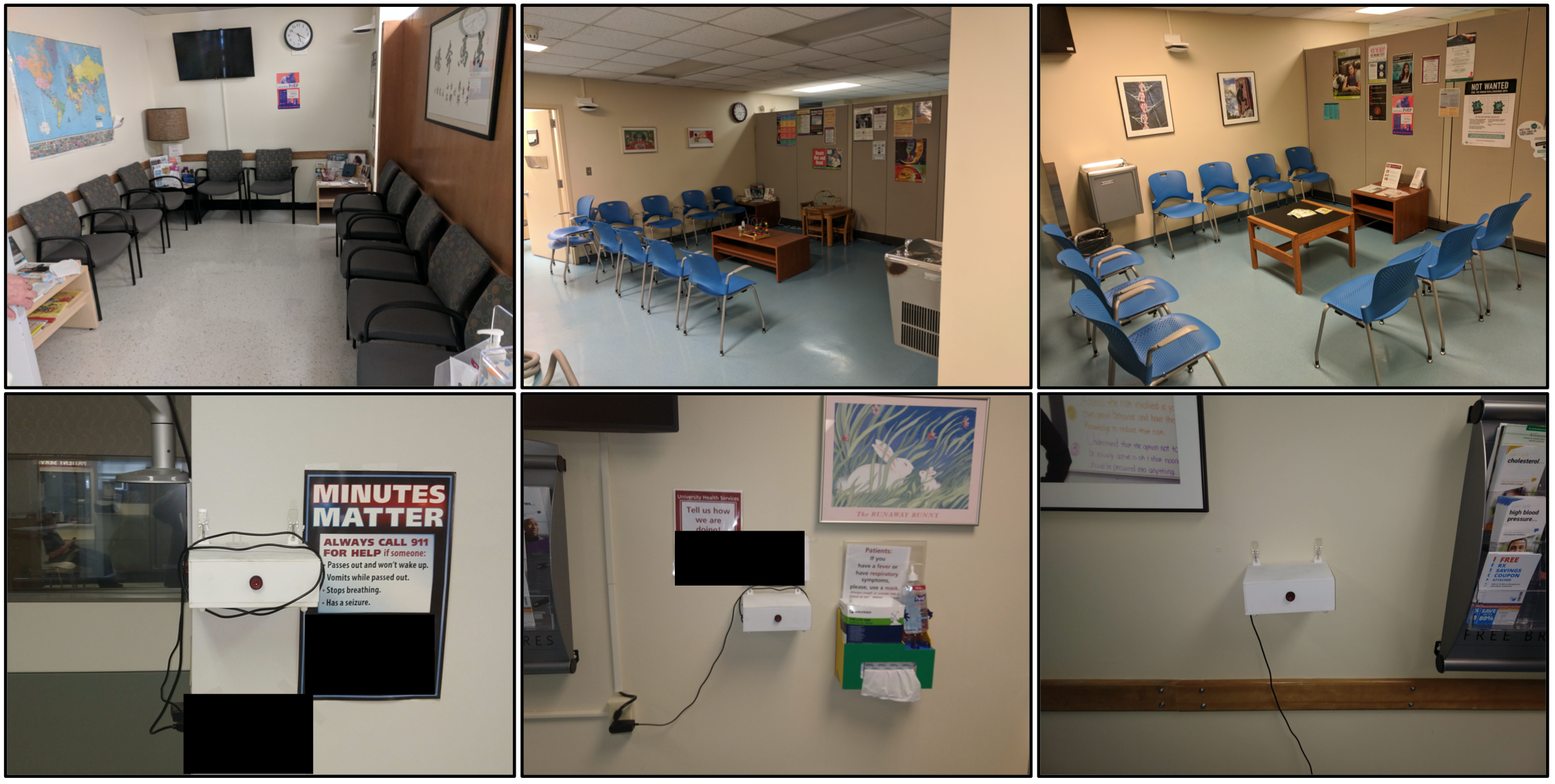}
  \caption{Waiting rooms and the placement of our sensor system}
  \label{fig:uhsDeployment}
\end{figure}

We deployed our device to collect data in a large public university's University Health Services (UHS) clinic building. The university has a student, faculty, and staff size of more than 30,000 people, and the UHS facility serves the basic health needs of all of the people in that university.

We deployed our device in the main waiting room for that university health services building. We started deploying our device on December 10, 2018 and finished our deployment on July 12, 2019. In the end, we collected around 4 million audio seconds of data from our deployment

\section{Audio Processing and Modeling}

\subsection{Multi-Channel Audio Processing}

We used single-channel audio captured by the Respeaker microphone by on-device beamforming \cite{URL:ReSpeaker}. This single channel audio stream was produced by applying 4-channel Generalized Cross-Correlation Phase Transformation (GCC-PHAT) algorithm \cite{gcc_phat}. In this algorithm, at first, the time delay of arrival (TDOA) is estimated by computing cross-correlation between different multi-channel audio streams captured by multiple microphones. With this information, audio from different channels are beam-formed to produce a single-channel audio stream.

\subsection{Handling Missing Audio Segments due to Presence of Speech}
\begin{figure}[t]
\begin{subfigure}{.5\textwidth}
  \centering
  \includegraphics[width=1\linewidth]{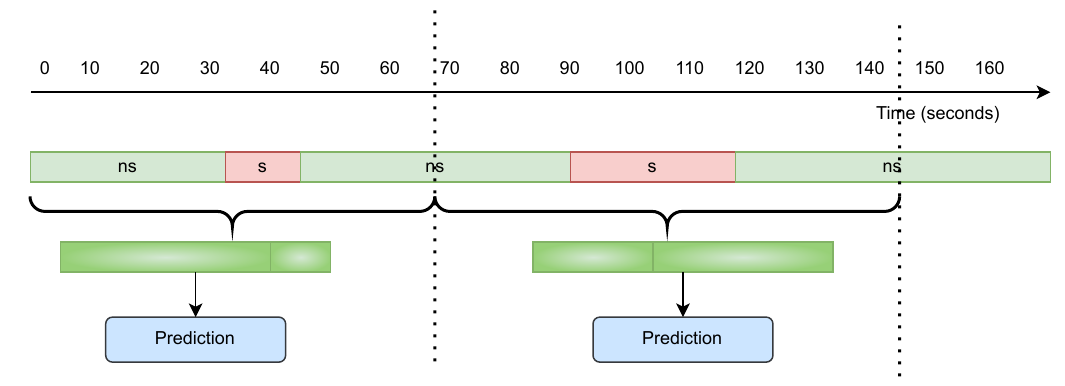}
  \caption{Scheme 1: construct spectogram using only non-speech segments (duration of 60 seconds) , speech probability information is not used}
  \label{fig:scheme1}
\end{subfigure}
\begin{subfigure}{.5\textwidth}
  \centering
  \includegraphics[width=1\linewidth]{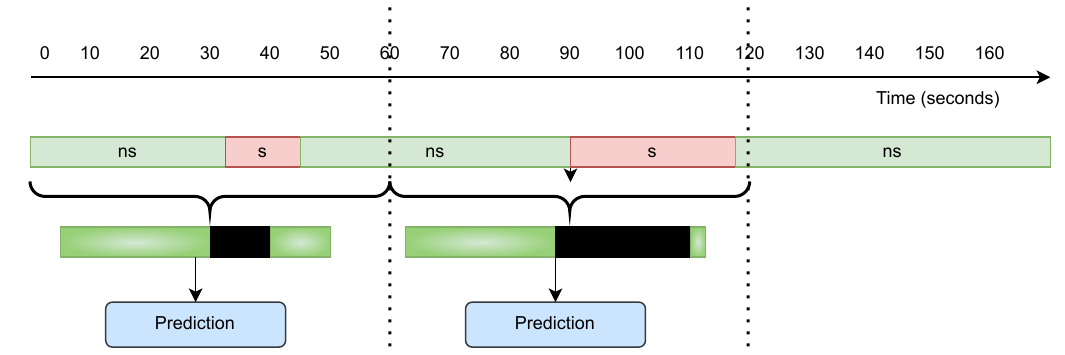}
  \caption{Scheme 2: construct spectogram using a cobination of non-speech segments and zero for segments detected as speech (duration of 60 seconds), speech probability is used as model input}
  \label{fig:scheme 2}
\end{subfigure}
\caption{Schemes for dealing with missing audio due to speech segments}
\label{fig:preprocess_scheme}
\end{figure}

Our system explored two different schemes for audio processing, as illustrated in Figure \ref{fig:preprocess_scheme}.

\subsubsection{Scheme 1} \label{sec:scheme1} 
In this scheme, the transformer model processes the audio stream data in 1-second chunks continuously. To remove speech-related information, a speech detection model is employed. Chunks identified as containing speech (with a probability greater than 0.5) are discarded. Non-speech audio chunks are accumulated until the count reaches 60. Afterwards, these 60 non-speech data points are fed into the transformer model as a sequence, enabling occupancy estimation.
 
\subsubsection{Scheme 2} \label{sec:scheme2}
In this scheme, a different approach was used for speech and non-speech chunks. For every second of audio data, if the chunk is identified as speech, the input spectrogram is replaced with all-zero values. In that case, only the speech probability obtained from the speech activity detection model is utilized as an input. On the other hand, for non-speech chunks, both the input spectrogram and the corresponding speech probability are used as inputs to the transformer model.

For example: Consider an audio stream from a hospital waiting room area with the following sequence: [n, n, s, n, s, n, n, n], where n denotes non-speech chunks, and s denotes audio chunks with speech.
Using Scheme 1, the chunks identified as speech (with probabilities above 0.5) are discarded, resulting in the following sequence of non-speech chunks: [n, n, n, n, n, n]. In comparison, using Scheme 2, this sequence will be converted to [n, n, z, n, z, n, n, n] where z represents all zero spectrograms. Besides these spectrograms, the speech probabilities are also fed into the model using this scheme.





\subsection{Audio Embedding}
To preprocess spectrograms for transformer models, we investigated two CNN encoder models: TRILL \cite{TRILL} and VGGish \cite{VGGish}.

\textbf{TRILL}: TRILL is a deep-learning model that can be used for extracting representation from non-semantic audio. Before feeding our data to the transformer model, for each second of audio, we extract 512-dimensional embedding for the audio snippet. This embedding is used as a representation of the non-speech audio snippet.

\textbf{VGGish}: VGGish is another deep-learning model that can be used to extract 128-dimensional representation from audio. However, unlike the TRILL model, we attached VGGish network to the input layers of our transformer and trained the whole model in an end to end fashion. This particular method ensures that besides optimizing the transformer model, we also optimize the VGGish model for generating suitable representation of audio.

In both cases, when the speech probabilities are used as input to the models (\ref{sec:scheme2}), we append the speech probabilities, and the end of these 512 (for TRILL) and 128 (for VGGish) dimensional representations before feeding the embedding representation sequence to the transformer model.

\begin{figure*}[ht!]
\begin{subfigure}{.65\textwidth}
  \centering
  \includegraphics[width=1\linewidth]{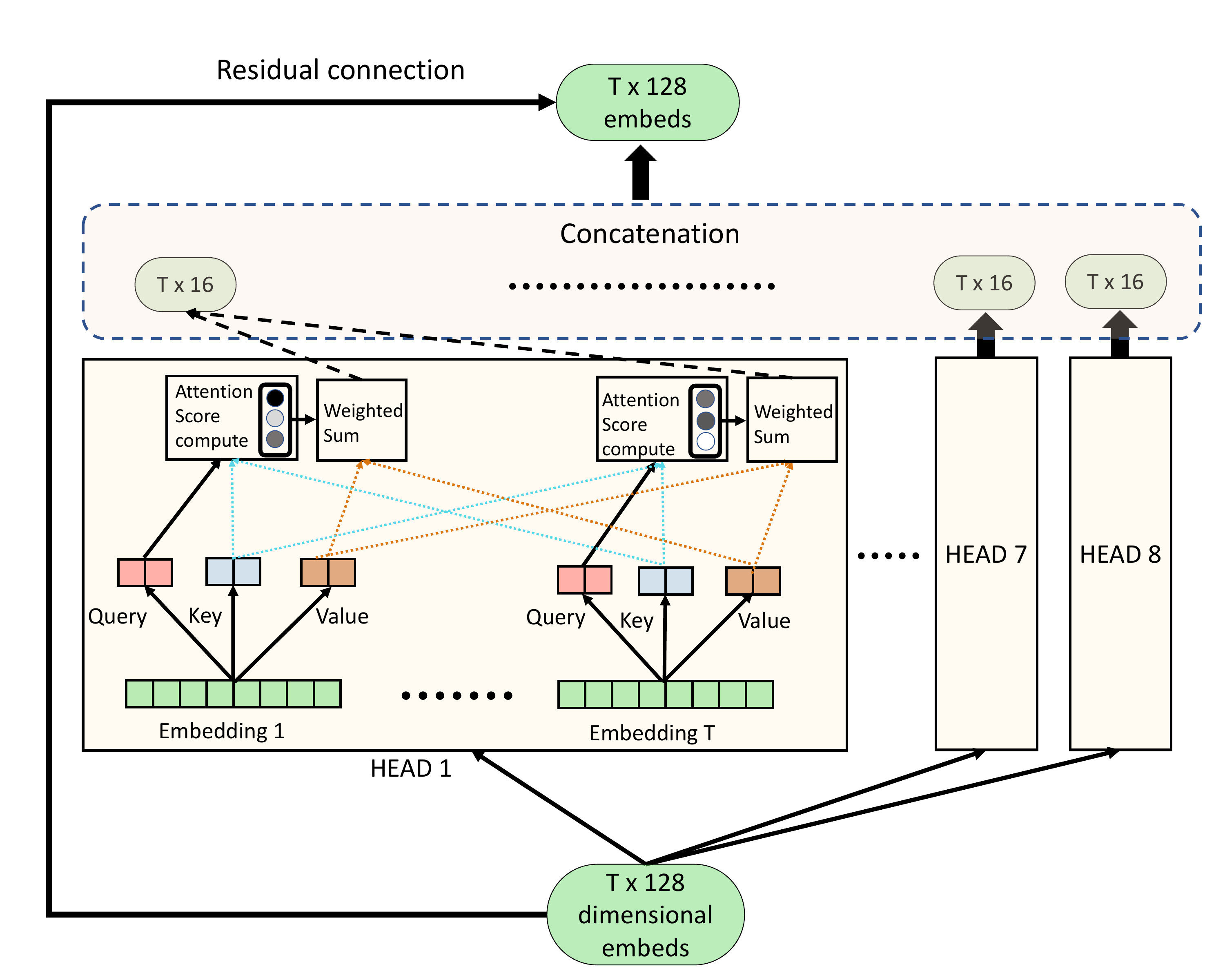}
  \caption{}
  \label{fig:fluTimeline}
\end{subfigure}
~
\begin{subfigure}{.25\textwidth}
  \centering
  \includegraphics[width=1\linewidth]{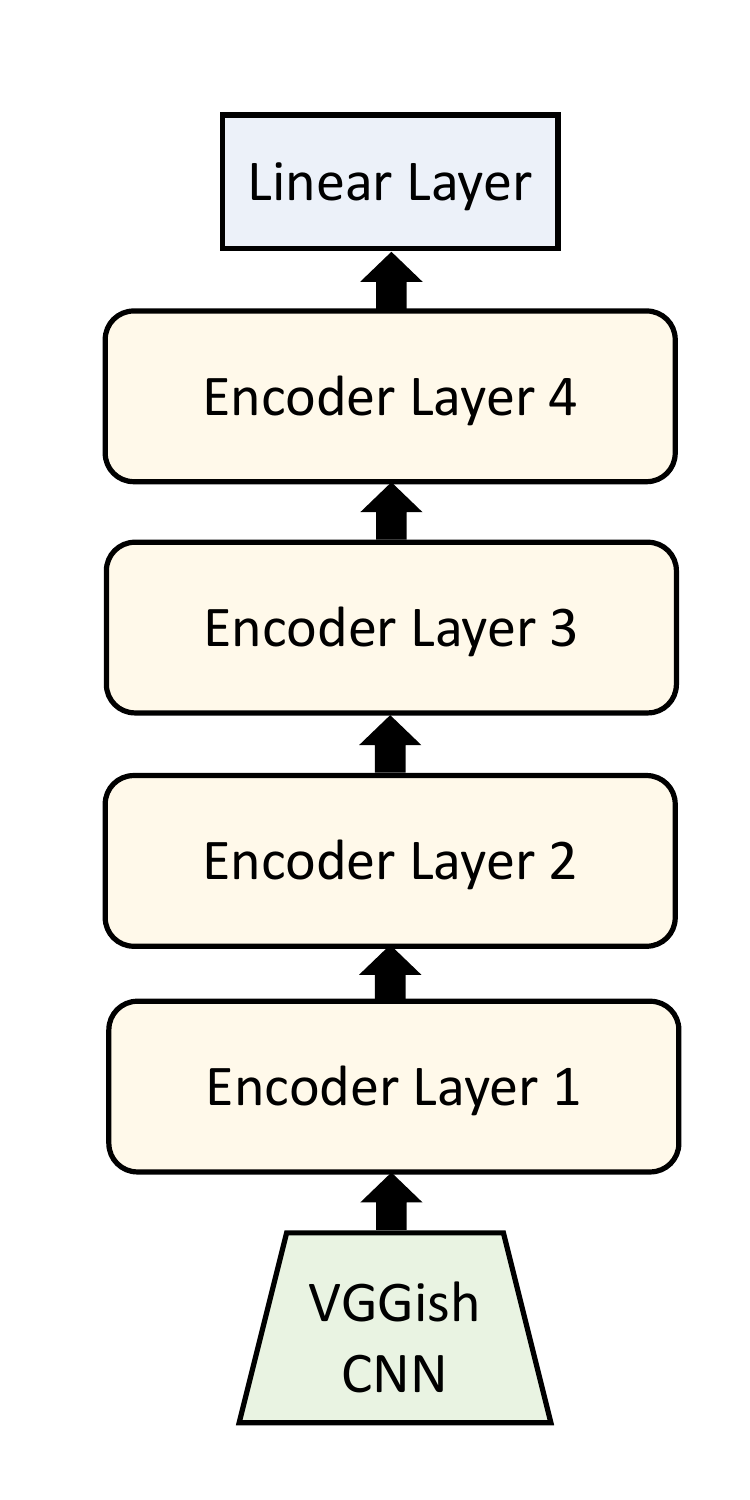}
  \caption{}
  \label{fig:fluTimeline}
\end{subfigure}
\caption{Overview of transformer model:  (a) Multi-head attention encoder (b) Our four-layer transformer model}
\end{figure*}

\subsection{Transformer model}

A transformer\cite{transformer_intro} is a time-series model that consists of several multi-head self-attention layers. A single head in a layer takes a sequence of  $T$ embeddings of dimension $d_{emb}$. For our case $d_{emb} = 128$ and projects each of the elements in the sequence to query ($Q$), key ($K$), and value ($V$) embeddings. For a single embedding $e$, the query, key, and value can be computed as follows: $Q = W_Q e$, $K = W_K e$, and $V = W_V e$. Here, $W_Q$, $W_K$, and $W_V$ are projection matrices. For our particular setting, we project 128-dimensional embeddings to a 16-dimensional space.




We calculate the values ($V$) by aggregating them using the attention score (computed using queries and keys)  as defined in the following equation:
$$\overline{V} = \sum_{j=1}^{T} \left( \frac{\exp(q_i^T k_j)}{\sum_{t=1}^{T} \exp(q_i^T k_t)} \right) V_j$$


From each head, for each embedding position, we produce these 16-dimensional embeddings. Finally, we concatenate the embeddings from 8 different heads, resulting in a 128-dimensional embedding output. The original 128-dimensional embedding values are added to this output via a residual connection\cite{ResNet}. The subsequent encoder layer takes this output after applying a ReLU activation function to it.

For our models, we used four such encoder layers. The output from the final layer goes through a linear layer to produce T outputs, which are our prediction values for occupancy in each second.

Intuitively, each transformer encoder layer model head finds the relationship between the input sequence elements passed from the previous layer. The value(V) represents an extracted feature from each of the elements of the sequence. These features are aggregated according to the self-attention score to create a richer set of representations for the sequence. With multiple heads, different sets of features are extracted. The final concatenated embeddings represent the aggregated version of all the features for that element in the sequence.  

Transformer-based models can take variable length sequence input similar to RNN and LSTM, and has several key advantages such as faster training and inference in comparison \cite{transformer_intro}. Other advantages such as the use of residual connections, not requiring backpropagation through time enable transformer models to be more efficient.

\subsection{Model Training}

\subsubsection{Ground Truth Collection}
In this section, we outline the process of ground truth data collection for our study. We extract ground truth data from the entry-exit data of the hospital waiting room, obtained on a daily basis. To accurately determine the number of individuals present in the waiting room at any given time, we created a people counting time series dataset for each day.

The time series dataset was constructed by initializing the patient count to 0 at the beginning of each day. The count was incremented as individuals entered the waiting room and reduced when they left. This approach provided a reliable ground truth dataset, allowing accurate assessment of the occupancy in the hospital waiting room at any given time.


\begin{table*}[!htbp]
\small
\centering
\caption{Mean Average Error (MAE), Root Mean Squared Error (RMSE), Correlation  Coefficient ($\rho$) between the occupancy prediction and ground truth for different models}
\begin{tabular}{|c|c|c|c|c|c|}
{\cellcolor[HTML]{C0C0C0}Model} & {\cellcolor[HTML]{C0C0C0}Modality} & {\cellcolor[HTML]{C0C0C0}Model Descripition} &  {\cellcolor[HTML]{C0C0C0} MAE} & {\cellcolor[HTML]{C0C0C0}RMSE} & {\cellcolor[HTML]{C0C0C0}$\rho$}\\ \hline
1 & Baseline(Average prediction) & Predict mean & 3.73 & 4.54 & 0.0 \\ \hline
2 & Thermal image  & Faster RCNN & 3.90 & 4.26  &  0.11\\ \hline
3 & Speech probabilities  & Random Forest & 3.23 & 4.16  &  0.34\\ \hline
4 & 1 channel(PHAT) Non-speech audio & Trill-embedding + Transformer & 2.18 & 3.00 & 0.74 \\ \hline
5 & 1 channel(PHAT) Non-speech audio & VGGish-CNN + Transformer & 2.23 & 3.00 & 0.73 \\ \hline
6 & Specch probability + 1 channel(PHAT) Non-speech audio & VGGish-CNN + Transformer & 2.46 & 3.21 & 0.69 \\ \hline
7 & Specch probability + 1 channel(PHAT) Non-speech audio & Trill + Transformer & 2.63 & 3.21 & 0.63 \\ \hline
\end{tabular}
\label{tab:main_results}
\end{table*}

\subsubsection{Model Setup and Training}
Our model is based on a 4-layer transformer encoder. We explore two different approaches for embedding: TRILL-based and the VGGish-based model.

For the TRILL-based model, we extract a 128-dimensional embedding and utilize it as input for the transformer model. On the other hand, for the VGGish-based model, we incorporate the VGGish model as the input for the transformer embeddings. We train this model in conjunction with our transformer model.

At each layer of the transformer, we employ 8 transformer heads to process the time series data. This allows for effective information processing at each layer. In the final layer, a linear layer converts the time series into a series of single numbers. As a result, for a 60-second audio input, we obtain an output of size $60 \times 1$. We define this output as representing the occupancy at any given time within the 60-second input.

To evaluate the performance of our model, we employ the Mean Squared Error (MSE) loss function, which measures the discrepancy between the ground truth and the output. During the training phase, we use the Adam Optimizer with a learning rate of $0.001$. We train the model for 30 epochs, allowing it to learn and optimize its performance over time.




\section{Results}
We develop a baseline model which predicts the mean occupancy, and train models using other data modalities, thermal image, and speech probability (no speech audio used) for every minute. We train the models with our collected waiting room data and leave out the last ten days of data for testing. We also perform cross validation by leaving out the data for entire months.

\subsection{Performance using different data modalities}

\begin{table*}[!htbp]
\centering
\caption{Results for different cross-validation months in terms of MAE (Mean Average Error), RMSE(Root Mean Squared Error), Correlation coefficient($\rho$)}
\label{tab:crossval_v2}
\begin{tabular}{@{}lrccccccccc@{}}
\toprule
\multicolumn{2}{l}{}                                                                            & \multicolumn{3}{c}{\textbf{MAE}} & \multicolumn{3}{c}{\textbf{RMSE}} & \multicolumn{3}{c}{\textbf{$\rho$}} \\ \cmidrule(lr){3-5} \cmidrule(lr){6-8} \cmidrule(lr){9-11}
\multicolumn{2}{r}{\textbf{Month}}                                                              & February    & March    & April   & February    & March    & April    & February    & March    & April   \\
\textbf{Model} & \textbf{Modality}                                                              &             &          &         &             &          &          &             &          &         \\ \midrule
1              & Baseline                                                                       & 3.18        & 2.68     & 3.93    & 4.02        & 3.21     & 4.74     & 0.00        & 0.00     & 0.00    \\
2              & Thermal image                                                                  & 3.34        & 2.94     & 3.57    & 4.39        & 3.44     & 4.49     & 0.14        & 0.19     & 0.14    \\
3              & Speech probability                                                             & 2.62        & 2.81     & 3.61    & 3.62        & 3.41     & 4.75     & 0.50        & 0.32     & 0.32    \\
-               & Speech probability + loudness      & 2.50        & 2.32     & 3.58    & 3.50        & 3.35     & 4.71     & 0.53        & 0.34     & 0.41    \\
4           & Non-speech Audio (TRILL)                                                               & 1.82        & 2.00     & 2.65    & 2.65        & 2.70     & 3.48     & 0.72        & 0.61     & 0.69    \\
5              & Non-speech Audio (VGGish pretrained) & 1.86        & 2.21     & 2.87    & 3.01        & 2.86     & 3.80     & 0.61        & 0.56     & 0.61    \\
6              & Speech probability + VGGish         & 1.95        & 2.13     & 2.82    & 2.90        & 2.77     & 3.88     & 0.70        & 0.63     & 0.68    \\
7              & Speech probability + Trill           & 1.95        & 2.14     & 2.96    & 2.90        & 2.80     & 3.89     & 0.70        & 0.54     & 0.61    \\ \bottomrule
\end{tabular}
\end{table*}

Table \ref{tab:main_results} shows the performance of different minute-level waiting room occupancy prediction models trained with different modalities including thermal camera, speech and non-speech. The first key takeaway is that the non-speech audio-based models outperform the thermal-based and speech-probability based models. While the thermal camera-based Faster RCNN model \cite{faster_rcnn} achieves a Person correlation coefficient ($\rho$) of 0.11 and a Mean Average Error (MAE) of 3.90, the non-speech audio based Transformer model with Trill embedding (scheme 1) achieves significantly better results with a $\rho$ of 0.74 and MAE of 2.18. Similarly, the non-speech audio based Transformer model with Trill embedding outperforms model that leverages speech probability information by a significant margin (i.e. $\rho$ of 0.34 for speech probabilities vs. $\rho$ of 0.74 for non-speech). Overall, this results clearly highlight that non-speech-based models contain information about waiting room occupancy which can be used to develop a privacy-aware occupancy system that solely relies on non-speech audio (and avoids analyzing any speech audio). This type of waiting room occupancy estimation system is particularly well-suited for clinical/hospital environment, as well as other public spaces that have higher level of privacy expectations.

We experimented with different schemes (section \ref{sec:scheme1}, \ref{sec:scheme2}) and embeddings as we develop our non-speech audio-based model. As can be seen in Table \ref{tab:main_results}, scheme 1 that uses 60 seconds of non-speech audio performs better than scheme 2 that leverages the extra information of speech probability (Note that only speech probability was used and no speech audio signal was used). For example, Model 4 with non-speech audio data achieves a $\rho$ of 0.74 and MAE of 2.18. Compared to this, Model 7 that uses both speech probability and non-speech audio as input to the same model architecture of Model 4 achieves a slightly lower performance with $\rho$ of 0.69 and MAE of 2.46. Overall, with the addition of speech probability in scheme 2, we get a slightly worse performance. It shows that the results of both of the schemes are comparable and after using non-speech sound adding speech probabilities does not contain that much information. 

\begin{figure}[!htbp]
\begin{subfigure}{.25\textwidth}
  \centering
  \includegraphics[width=1\linewidth]{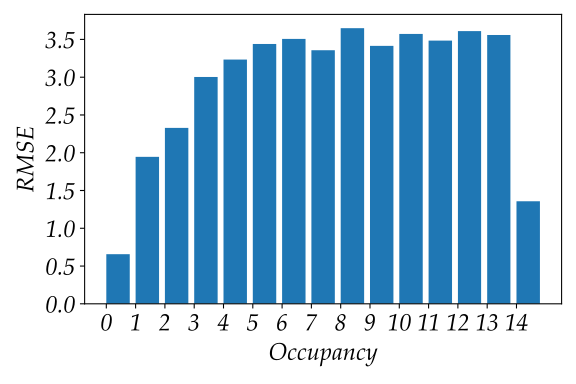}
  \caption{}
  \label{fig:trill_emb_rmse_cv}
\end{subfigure}
~
\begin{subfigure}{.25\textwidth}
  \centering
  \includegraphics[width=1\linewidth]{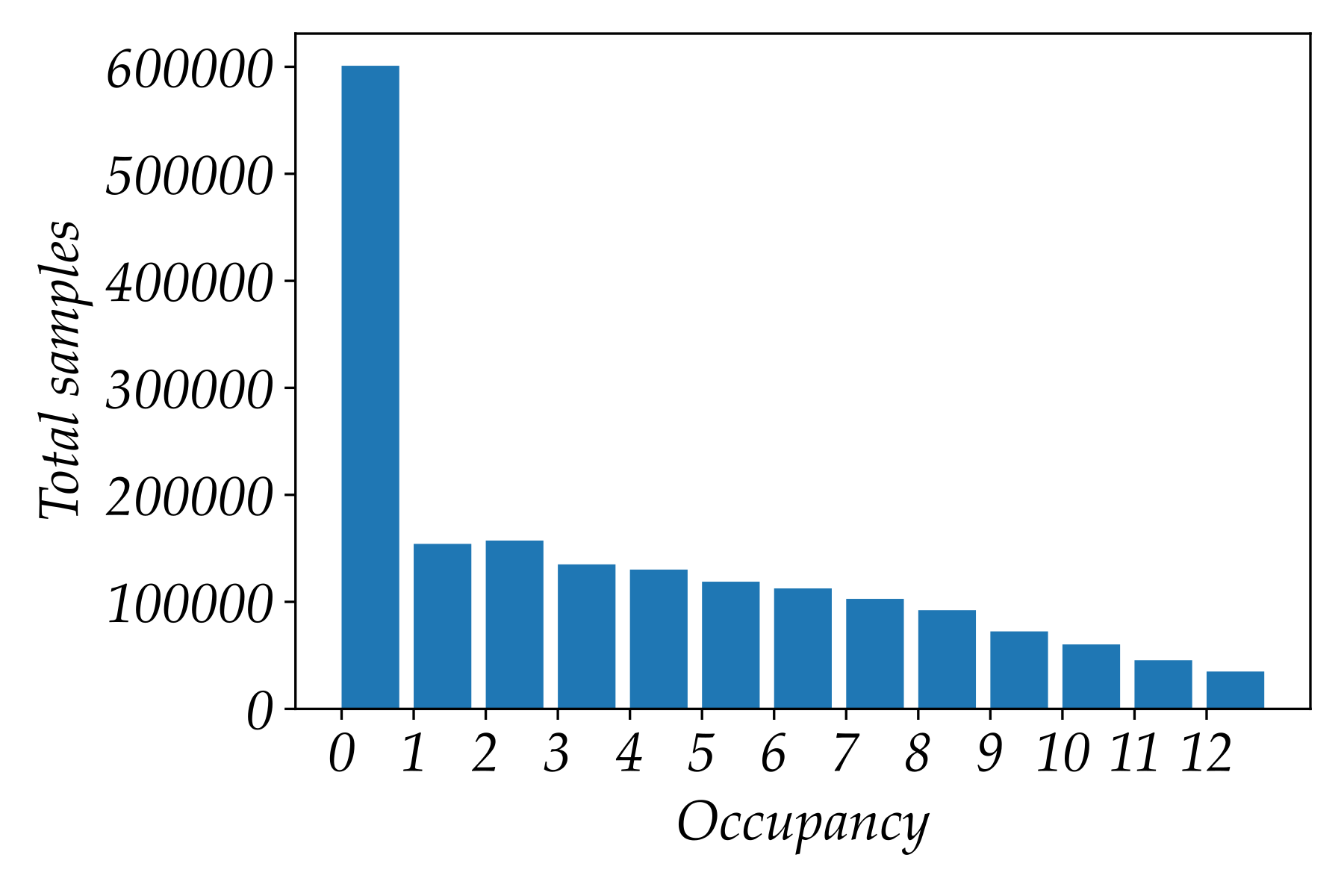}
  \caption{}
  \label{fig:samps_by_occ}
\end{subfigure}
\caption{(a) Plot showing average RMSE by occupancy (b) Total samples(in seconds) in the dataset by occupancy}
\label{fig:performance_by_occ}
\end{figure}

\begin{figure}[!htbp]
\begin{subfigure}{.25\textwidth}
  \centering
  \includegraphics[width=1\linewidth]{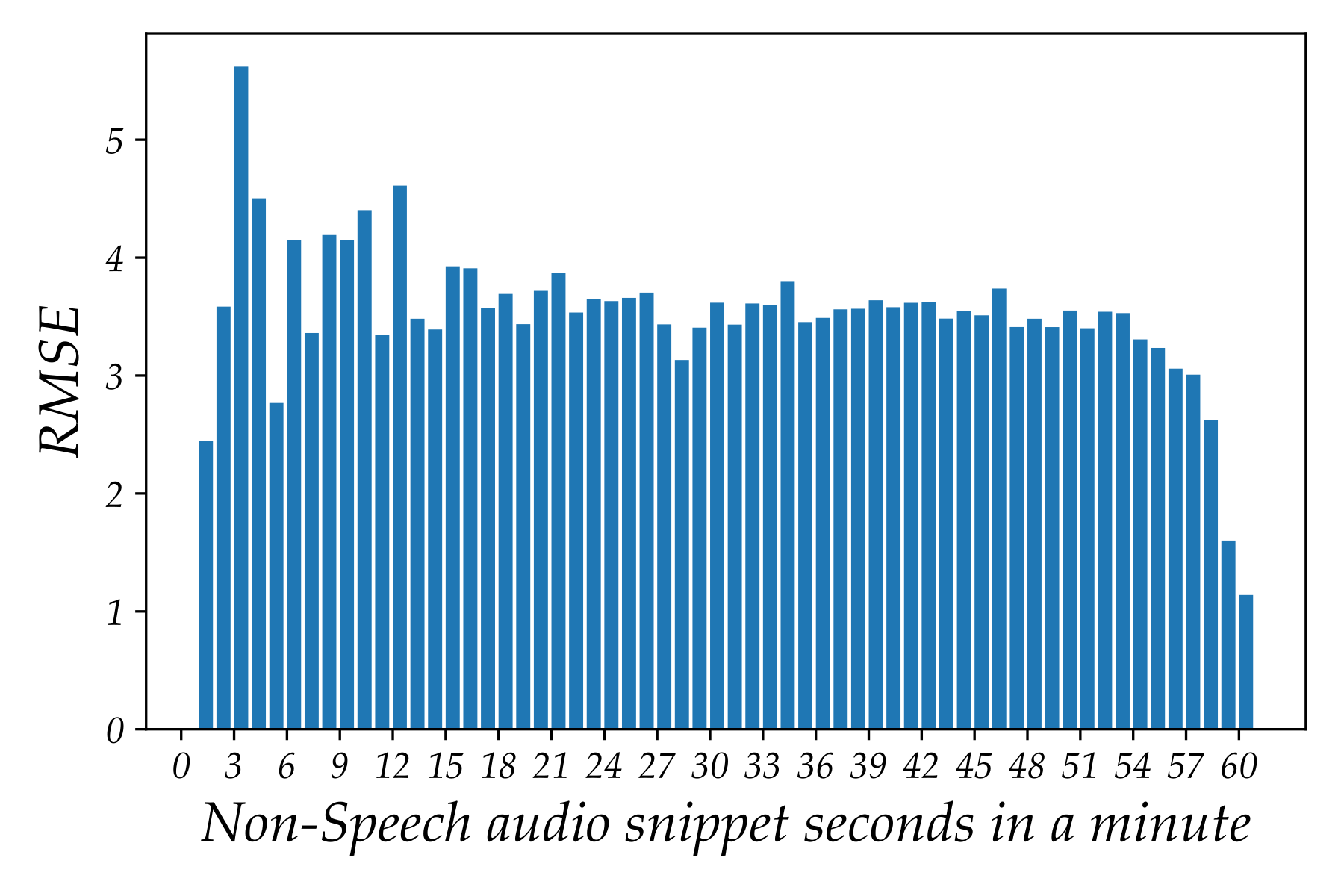}
  \caption{}
  \label{fig:trill_emb_rmse_cv}
\end{subfigure}
~
\begin{subfigure}{.25\textwidth}
  \centering
  \includegraphics[width=1\linewidth]{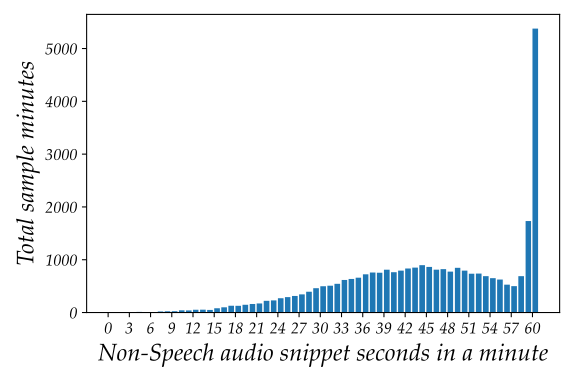}
  \caption{}
  \label{fig:samps_by_occ}
\end{subfigure}
\caption{(a) Plot showing average RMSE by occupancy (b) Total samples(in seconds) in the dataset by occupancy}
\label{fig:non_speech_in_minute}
\end{figure}

In Figure \ref{fig:non_speech_in_minute}, we can observe that the majority of the collected real life data consists of non-speech snippets for the entire minute. Having a substantial amount of non-speech data available enhances the performance of our models. Conversely, when the amount of non-speech snippets is lower, the performance remains relatively unchanged and does not drop. In Figure \ref{fig:performance_by_occ}, we observe that we have more data for lower occupancy setting and our model has lowest error for such scenarios. The performance of our model decreases slightly for higher occupancy setting by negligible amount.

\subsection{Time scale aggregation}
We aggregate the predictions from our minute-level occupancy models and compare the correlation coefficient with the ground truth data in Table \ref{tab:multi_timescale}. It can be seen that $\rho$ improves if a longer aggregation window is used for all modalities. However, the key takeaway is that non-speech audio-based model outperforms the models using other modalities by substantial margins for time aggregated results ($\rho \geq 0.90$ for non speech audio compared to $\rho < 0.25$ using thermal image and $\rho < 0.50$ using speech probability).



\begin{table}[ht!]
\centering
\caption{Correlation coefficient of average aggregation results for different modalities for different timescales}
\begin{tabular}{cccc}
\toprule
Data &   30 min $\rho$ & 1 hour $\rho$ & 2 hour $\rho$\\ \midrule
Thermal image & 0.23 & 0.24  &  0.24\\ 
Speech probability & 0.44 & 0.47 & 0.48 \\ 
Non-speech Audio  & 0.90 & 0.91  & 0.93 \\ 
\bottomrule
\end{tabular}
\label{tab:multi_timescale}
\end{table}

\subsection{Cross validation}

We also conducted leave-one-month-out cross validation, where the entire data from a specific month was used as test data. The results are shown in Table \ref{tab:crossval_v2}. Consistent with the results in the previous section, the cross-validation results demonstrate superior performance using the non-speech modality. Using scheme 1, non-speech audio using TRILL embedding obtains the best performance for all three of the months compared to using other modalities and audio processing schemes in terms of MAE, RMSE and $\rho$. Using scheme 2, we also obtain better performance than using the other modalities.

\begin{figure}[!htbp]
\begin{subfigure}{.25\textwidth}
  \centering
  \includegraphics[width=1\linewidth]{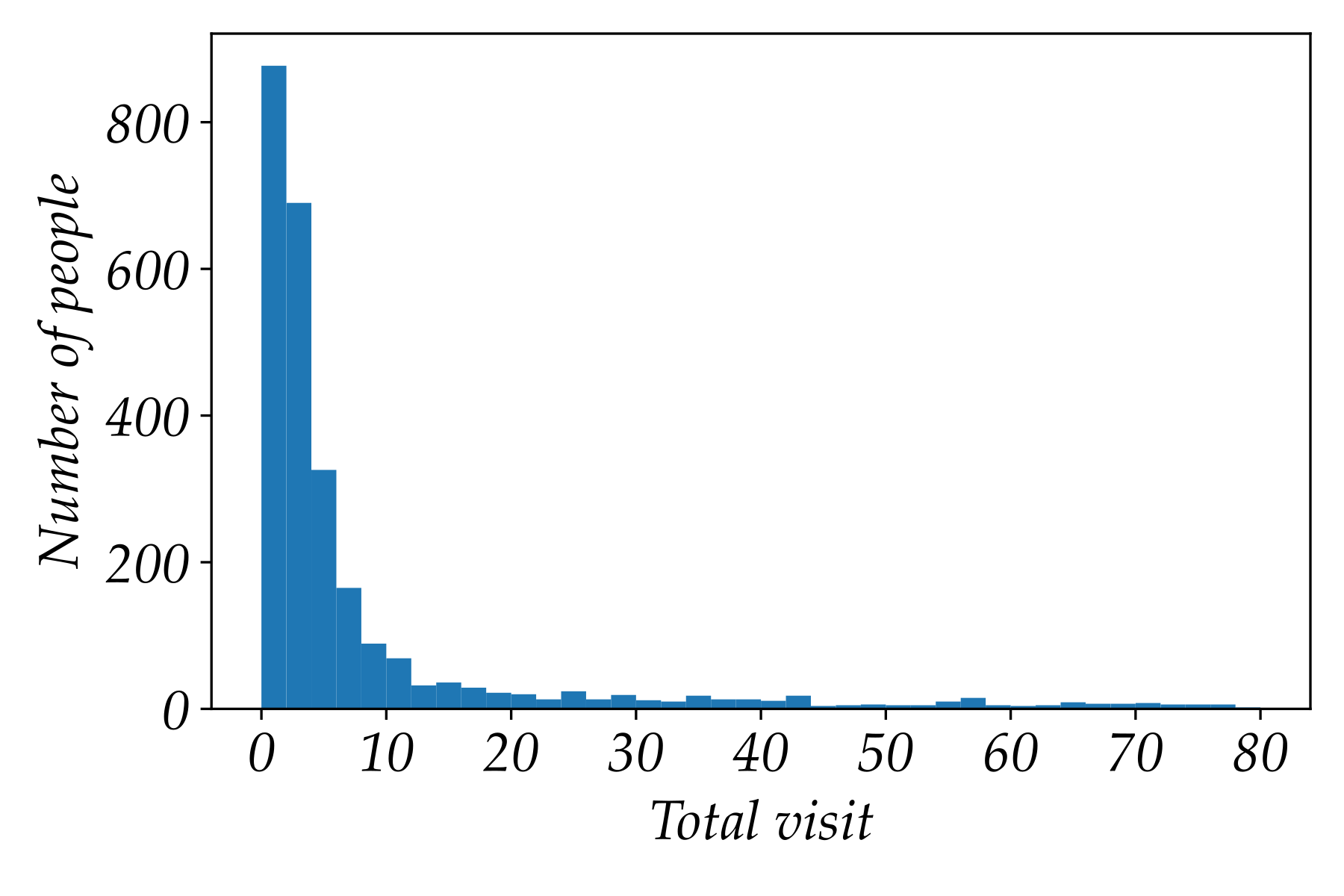}
  \caption{}
  \label{fig:trill_emb_rmse_cv}
\end{subfigure}
~
\begin{subfigure}{.25\textwidth}
  \centering
  \includegraphics[width=1\linewidth]{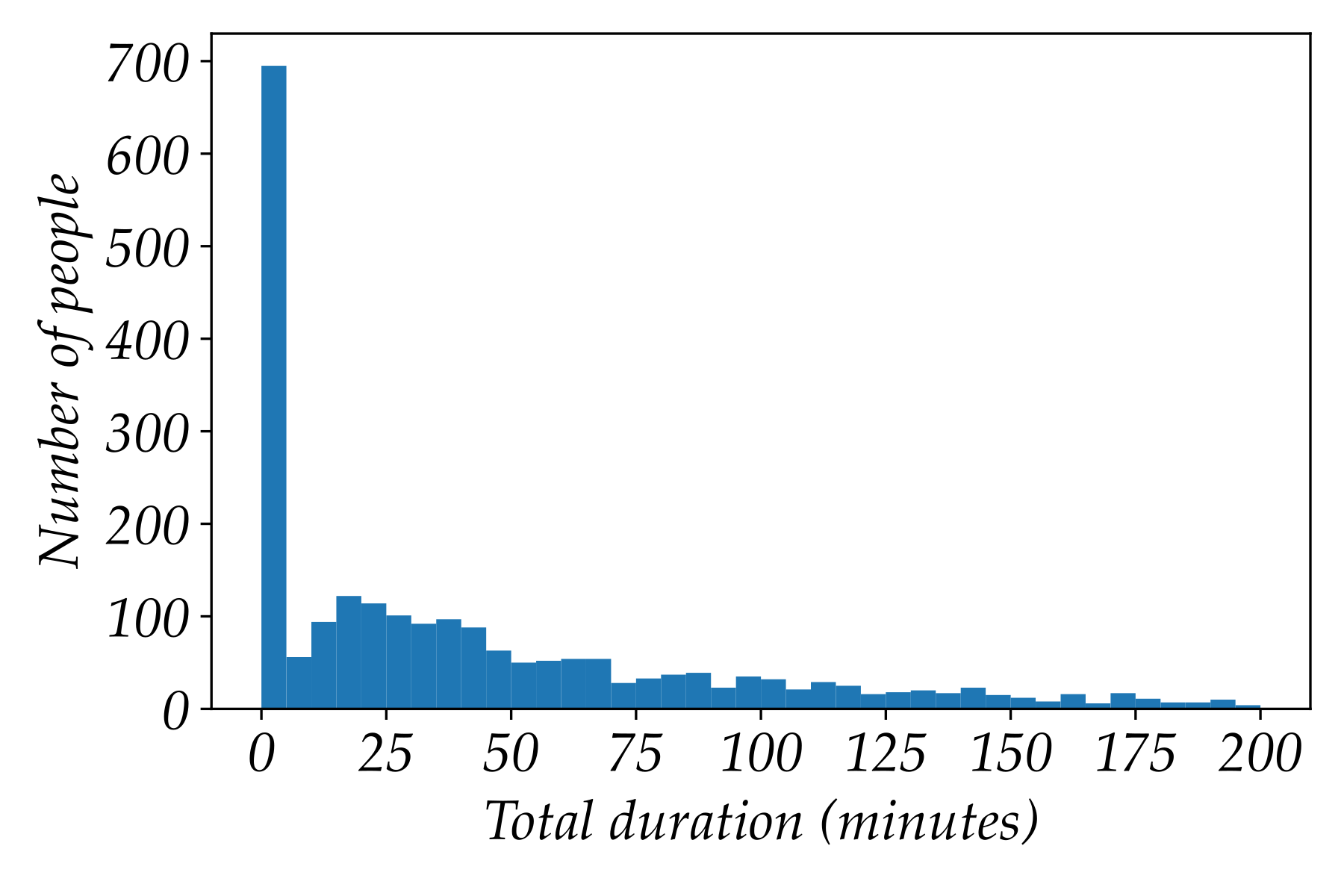}
  \caption{}
  \label{fig:samps_by_occ}
\end{subfigure}
\caption{(a) Plot showing distribution of visits by patients (b) Plot showing distribution of total duration of visits by patients}
\label{fig:patient-visit-distribution}
\end{figure}

\section{Privacy preserving Inference}

\subsection{Information linkage attacks and data privacy}
Linkage attacks, exemplified by the case of Netflix \cite{Netflix_wired}, pose a significant security threat by linking new data with existing public data, potentially compromising individuals' privacy. To combat this risk, differential privacy has emerged as a powerful framework. It provides a rigorous mathematical guarantee to protect individual privacy while enabling meaningful analysis of sensitive data. By injecting controlled noise into data analysis, differential privacy prevents specific information extraction and safeguards privacy against linkage attacks. It strikes a balance between privacy preservation and data utility, allowing valuable insights while upholding individuals' privacy rights. Adopting differential privacy techniques is crucial to address privacy concerns and ensure responsible use of data in the face of evolving security challenges.

As our system processes non-speech audio data, we can implement differential privacy by adding noise to the embeddings generated by our CNN model (which is then later fed into the transformer model). As the transformer model can be computationally heavy, while the CNN features can be computed in real-time on-device, this methodology ensures that we only send differentially private information to the backend for further processing.

\subsection{Implementation of Differential Privacy guarantees for our system} \label{sec:diff_priv_intro}

\newcommand{\Lone}[1]{\left\lVert#1\right\rVert_{1}}
\newcommand{\Ltwo}[1]{\left\lVert#1\right\rVert_{2}}
To ensure our system outputs data with differential privacy requirements, we use Laplace mechanism to add noise to the output of our CNN based audio embeddging generation system \cite{diff_privacy_book}. To do that, at first we define the sensitivity of any function $f$ as  
$$\Delta f = \max_{\substack{x, y \in \mathbb{N}^{k} \\ {\Lone{x - y} = 1}}} \Lone{f(x) - f(y)}$$

where $f: \mathbb{N}^{|{\mathcal{X}}|} \rightarrow \mathbb{R}^{k}$. With this, definition, one can prove that, with the Laplace mechanism on the output of $f(x)$ defined as:

$$\mathcal{M}_{L}(f(x)) = f(x) + (Y_{1} , Y_{2}... Y_{k})$$
where each of $Y_i$ are i.i.d variables drawn from Laplace distribution of $L(\frac{\Delta f}{\epsilon}) = \frac{\epsilon}{2\Delta f}\exp(-\frac{|t|\epsilon}{\Delta f})$; is an ($\epsilon$ ,0) differential private algorithm.

Based on this analysis; for our modeling scenario if we extract $d$ dimensional embeddings from $t$ timesteps (from $t$ seconds of audio clips); our embedding generation function will produce $t \times d$ dimensional output. If we clip the  output embedding for each $i$th second as:
$$\tilde{e_i} = \frac{e_i}{\max(1, \frac{\Lone{e_i}}{C})}$$
Where $C$ is the clipping parameter; which is a hyperparameter.
From this formulation we can conclude that: $$\Lone{\tilde{e}} \leq C$$ 
So, for two $t \times d$ embedding sets where in one embedding set the embedding in some particular second $i$ is hidden (which means not using that particular second of audio); say $e_{-i}$ by setting those positions with zeros compared to another embedding set where the embedding from $i$th second is present ($e$), we can express the relation as:
$e = e_{-i} + e_{i} $

where $e_{i}$ a zero vector of size $t \times d$ except from positions $(i - 1) * d + 1$ to $i * d$; where it contains clipped embeddings from $i$th second. 

So from the above, we can conclude:
$$\Lone{e - e_{-i}} = \Lone{e_{i}} \leq C$$
Thus the embedding space has a sensitivity of $C$. From that, we can conclude for per clip, $(\epsilon, 0)$ differential privacy, all we need to do is perturb the embeddings with Laplace noise sampled from $L(\frac{C}{\epsilon})$ and add it to our embedding. This procedure ensures adding each second of audio and sending it to a remote backend (where the transformer-based occupancy detection model might run) has an at-most $1 + \epsilon$ diminishing effect return on any future utility \cite{diff_privacy_book}.

In this procedure, if a user decides to stay in the place $T$ seconds, the total differential privacy loss will be $\epsilon T$ because of the composition theorem of differential privacy mentioned at \cite{diff_privacy_book}.

\subsection{Experimental Results for our differential privacy formulation}

To make our models compatible with differential privacy, we first trained our model where CNN embeddings were clipped as described in \ref{sec:diff_priv_intro}. We also added noise according to the Laplace mechanism during the training to make the model prediction more robust with differential privacy noise.
Table \ref{tab:diff_privacy} shows the effect on the occupancy estimation algorithm for different $\epsilon$ differential privacy targets:  
\begin{table}[!htbp]
\small
\centering
\caption{Results for epsilon-delta differential privacy settings}
\begin{tabular}{|c|c|c|c|c|}
{\cellcolor[HTML]{C0C0C0} Mechanism} &  
{\cellcolor[HTML]{C0C0C0} $\epsilon$} & 
{\cellcolor[HTML]{C0C0C0} MAE} & {\cellcolor[HTML]{C0C0C0} RMSE} & {\cellcolor[HTML]{C0C0C0}$\rho$}\\ \hline
Laplace & 5  & 2.74 & 3.80 & 0.69 \\ \hline
Laplace & 2 & 2.87 & 3.97 & 0.66 \\ \hline
Laplace & 1 & 3.03 & 4.18 & 0.62 \\ \hline
Laplace & 0.5& 3.18 & 4.34 & 0.55 \\ \hline
Laplace & 0.25 & 3.60 & 4.80 & 0.41 \\ \hline
Laplace & 0.10 & 4.12 & 5.31 & 0.19 \\ \hline
\end{tabular}
\label{tab:diff_privacy}
\end{table}%

From table \ref{tab:diff_privacy}, we can see that our algorithm even performs well for $\epsilon$ values as small as 1. For the other values > 0.5, we can see that the MAE remains consistently low compared to the baseline.  With these small $\epsilon$ values, we can ensure that using our methodology reduces the possibility of future linkage attacks for each additional audio snippet we use for occupancy detection with our model.

%% file: discussion.tex
\section{Discussion and Conclusion}
In this work, we demonstrate the feasibility of estimating occupancy using non-speech audio, presenting superior results compared to other modalities, including thermal cameras. Compared to techniques that aim to preserve privacy while still using speech data, our results show that non-speech audio contains sufficient information to estimate occupancy reliably. Our non-speech-based method not only protects speech content but also renders speaker identification extremely difficult. Our analysis with differential privacy techniques shows that we can provide additional privacy guarantees on top of the privacy preserving data modality that we use.

Our approach incorporates a transformer-based model which has proven to be effective across various tasks and domains. By leveraging this non-speech based method, we enable the application of such models even in situations where deploying such models on edge-devices might not be feasible.

One notable advantage of employing non-speech audio lies in the richness of information that is contained, resulting in more reliable occupancy estimates. While alternative modalities, such as $\textit{CO}_{2}$ sensor and PIR sensors,may be effective in detecting the presence of humans, they often fall short in providing accurate occupancy information, as evident from the existing literature.

Furthermore, our analysis incorporates differential privacy techniques, adding an additional layer of privacy guarantees to the already privacy-preserving data modality we use. By incorporating differential privacy safeguards, we have not only ensured the present security of our system but also fortified it against potential future linkage attacks. This robust approach guarantees the protection of sensitive data, reinforcing the trustworthiness of our system for long-term use.

One of the limitations of our current approach is that we have only deployed our sensors in a hospital waiting room. To confirm the general applicability of our method, further investigations and evaluations in other scenarios may be required. We argue that our method is highly suitable for similar indoor environments with higher privacy expectation.

Our findings have a significant impact on the design and implementation of smart technology for homes, public spaces, and workplaces. Our approach could help to instill public trust by providing privacy protection while providing valuable insight beyond occupancy such as syndromic surveillance \cite{al2020flusense}. 

To conclude, our study demonstrates the feasibility and advantages of estimating occupancy using non-speech audio. Our approach surpasses other modalities in terms of privacy preservation and accuracy of occupancy estimation.